\documentclass[useAMS,usenatbib,referee]{mn2e}

\usepackage{epsfig}
\usepackage{times}
\usepackage{journalnames}

\def\MSun{{\rm\thinspace M_{\odot}}}

\def\apgt{\ {\raise-.5ex\hbox{$\buildrel>\over\sim$}}\ }
\def\aplt{\ {\raise-.5ex\hbox{$\buildrel<\over\sim$}}\ }

\title[]{The formation of planets in circumbinary disks}
\author[F.I. Pelupessy and S. Portegies Zwart]{
 F. I. Pelupessy$^1$\thanks{E-mail: pelupes@strw.leidenuniv.nl} and 
 S. Portegies Zwart$^1$ \\
 $^1$Leiden Observatory, Leiden University, PO Box 9513,
 2300 RA, Leiden, The Netherlands \\ \\
}

\begin{document}

\date{}

\maketitle

\begin{abstract}

We examine the formation of planets around binary stars in light of
the recently discovered systems Kepler 16, 34 and 35. We conduct
hydrodynamical simulations of self gravitating disks around binary
systems. The selected binary and disk parameters are chosen consistent with
observed systems. The disks are evolved until they settle in a
quasi-equilibrium and the resulting systems are compared with the
parameters of Kepler 16, 34 and 35. We find a close correspondence of
the peak density at the inner disk gap and the orbit of the observed
planets. We conclude, based on our simulations, that the orbits of the
observed Kepler planets are determined by the size of the inner
disk gap which for these systems results from the binary driving. 
This mediates planet formation either through the density enhancement 
or through planetary trapping
at the density gradient inversion in the inner disk. For all three systems the current eccentricity of 
the planetary orbit is less than the disk eccentricity in the simulations. This,
together with the long term stability of the orbits argues 
against in situ formation (e.g. a direct
collapse scenario of the material in the ring). Conducting additional
simulations of systems with a wider range of parameters (taken from a
survey of eclipsing binaries), we find that the planet semi-major axis
and binary eccentricity in such a scenario should be tightly
correlated providing an observational test of this formation
mechanism.

\end{abstract}


\section{Introduction}

The recent discovery of a population of planets in orbit around binary
stars \citep{Doyle2011, Welsh2012, Orosz2012a, Orosz2012b} adds an extra layer of complexity
to planet formation theories, which are already struggling to explain
the wide variety of systems uncovered by the current explosion of
planet discoveries \citep{Baraffe2010}. Around a single star the time
required to grow a giant planet is expected to be at least comparable,
and probably exceeds, the lifetime of the protoplanetary disk \citep{Yin2002}
and is subject to a number of bottlenecks. After the formation of the
 gaseous protoplanetary disk \citep{Wyatt2008} giant planets form through a
number of subsequent phases each of which is expected to last at least
a million years. The first planetesimal seeds  are thought to form from the 
dense dustlayer in the midplane of the disk, either through gravitational 
instabilities directly\citep{Michikoshi2010}, or by the fast collapse 
\citep{Johansen2007} and/or slow sedimentation of aggregations
that form in a turbulent gas velocity field\citep{Cuzzi2008}.  The
subsequent growth of these planetesimals to rocky proto planet is
mediated by oligarchic accretion \citep{Ida1993,Kokubo1998}, followed
by the accretion of a giant envelope by capturing gas from the
remaining disk \citep{Fortier2009}.  The growth of the rock and the
accretion of a gaseous envelope are not necessarily the most time
consuming phases, and each of these can be realized in a few Myr
\citep {Fortier2009}.

The most difficult part in the planet formation process appears to be 
the earlier phase in which the disk develops a highly contrasted density 
structure \citep{Michikoshi2010}, that eventually collapses under its 
own gravity to form the first planetesimals. In a disk around a single 
star the gravitational instability grows slower than the time scale for 
the dust to settle in a thin disk \citep{Yamoto2006}. Forming planets in 
such an environment is a non-trivial endeavour, and it may be mediated 
by turbulence in the dust disk \citep{Johansen2007, Cuzzi2008}.
 When the inner object is a binary rather than a single star, the
orbital motion excites perturbations in the disk. These perturbations 
may hinder the growth of planetesimals by increasing their orbital 
eccentricities and hence their relative impact velocities 
\citep[e.g.][]{Marzari2008, Paardekooper2012}.

In addition, the core accretion theory suffers from the problem that
proto-planetary cores are expected to migrate inwards before they have
a chance to grow in size.  This planet migration model would
naturally explain the presence of planets in close orbits around
their parent stars \citep{Ward1997,Tanaka2002}, but have difficulty to
reproduce the observed wide orbits.  On the other hand, the direct
formation scenario for gas giants via the collapse of disc
instabilities, can only explain a limited set of planets in remote
orbits \citep{Boley2009,Janson2012}.

Recently the concept of 'planet traps' was introduced to overcome the 
problems caused by (type I) planet migration: it turns out that in 
regions of positive (increasing density with radius) radial density 
slope the net effect of the corotation and Lindblad resonance torques 
flips sign \citep{Masset2006}. This leads to a (possible) pile up of 
proto-planetary cores at the radii of positive slope \citep[Note that 
additionally, these regions can have an inward pressure force leading to 
super Keplerian gas motions. As a consequence the gas drag on 
planetesimals also flips sign here and these regions are expected to 
show enhanced density for this reason;][]{Masset2006}.

The details of this process remain uncertain. The principal problem is 
that the actual radius where planets are trapped is ill constrained for 
disks around single stars \citep{Morbidelli2008}. If the trapping 
happens at the transition between the inner turbulent region and the 
'dead zone' it is uncertain at what distance and at what time the 
process is operating.  The distance for this process could be as close 
to the parent star as $\sim 0.1$\,AU \citep{Ilgner2006}, up to a 
distance of $\sim 5$\,AU \citep{Turner2007}, and the effective distance 
may even move with time. This is not necessarily a problem for this 
mechanism, but it makes it difficult to identify whether or not the 
planetary trapping process has taken place.

Circumbinary protoplanetary discs on the other hand have a definite
scale and will have a more precisely defined region where the density
gradient is positive. Circumbinary discs therefore do not suffer
from the aforementioned finetuning problem. The inner regions are
cleared out by the binary torque to radii $R> 2.3 a$
\citep{Artymowicz1994, Artymowicz1996}. The precise location of the
inner boundary is a weak function of the system parameters (semi-major
axis $a$, eccentricity $e$ and mass ratio $q$).

Hence circumbinary systems provide a test bed for planet formation
theories. The recent detection of circumbinary planets around Kepler
16, 34 and 35\footnote{ during revision of the manuscript additional circumbinary planet 
systems, Kepler 38 and 47, where published.} provides well constrained systems and a unique
opportunity to test planet formation theories. The binaries themselves
have sufficiently large separations that they are not expected to have
experienced major secular evolution of their orbital elements through
tidal interactions \citep{Devor2008}.  While it cannot be ruled out
that the planets themselves have migrated after their formation, the
orbital stability and the co-alignment of the angular momenta of the
orbits and the stellar spins \citep{Winn2011} suggest that these
systems have retained their orbital parameters since birth
\citep{Doyle2011}.

The formation of planets around binaries may be quite common, and their 
apparent lack can easily be the result of biases in the techniques used 
to search for planets.  The majority of solar type stars are in binary 
systems \cite{Duquennoy1991,Ghez1993}, and each may have formed with a 
circum binary disk. However, the lifetime of disks around close 
binaries may present a constraint for planet formation: \citep
{Kraus2012} found for the Taurus Auriga star forming region that $2/3$ 
of close binaries (semi-major axis $a \aplt 40$\,AU ) have dispersed 
their disk by 1 Myr, whereas 80\%--90\% of wide binaries and single 
stars retain their disk for more than 2-3 Myr. Even so, 1/3 of close 
binaries in the Kraus et al. study retain a disk for up to 10\, Myr, so 
circumbinary planet formation does not seem to be precluded. If 
survival of these disks allows for the formation of circumbinary 
planets in a seizable fraction of the binary population, which seems to 
be  confirmed by the detection of Kepler 16, 34 and 35 \citep{Welsh2012}, 
the prospects for detection by transit events are best in 
eclipsing binary systems \citep[e.g.][]{Devor2008} (assuming that 
circumbinary disks preferentially align with the binary, which is the 
case for the Kepler systems.) .

In this manuscript we study the formation of perturbations in 
hydrodynamic circumbinary disks and study the evolution of the density 
structure. We aim our study at the recently discovered systems Kepler 
16, 34 and 35, with the aim of making a detailed comparisons with the 
observed planetary parameters and assess whether these provide 
constraints on the formation channel of these systems. 

\section{Methods}
\label{Sec:Methods}

We simulate the structural evolution of a gaseous circumbinary disk 
around a binary star system. The numerical framework consists of a 
variety of ingredients brought together in the Astrophysics Multipurpose 
Software Environment (AMUSE) \citep[][]{PortegiesZwart2009,PortegiesZwart2012}. \footnote{The framework and the source 
codes of the scripts which were used to run the simulations presented 
here can be downloaded from \texttt {www.amusecode.org}.} 

The AMUSE package combines well tested simulation codes into a
software suite which can be used to perform individual tasks, or
reassemble the parts into a new application that combines a wide
variety of solvers. The interfaces of codes within a common domain are
designed to be as homogeneous as possible. The AMUSE application
consists of a user script, written in python, that controls the
community modules.  The user script specifies the initial conditions,
manages the calling sequence and data flow between the community
modules, controls the runtime error handling, checks for energy
conservation and other runtime diagnostics and performs a primary
analysis on the raw simulation data.  In the AMUSE philosophy we use
Python as a glue language to bind the community modules together.  The
relatively low speed of this high-level language is not an issue,
because most of the work is done in the community codes. 

For the simulations presented in this manuscript we co-evolved binary 
systems and a gaseous disc. The binary solver employed was a 
Kepler solver in universal variables \cite[e.g.][]{Bate1971}. The 
gaseous disk is simulated using a self-gravitating smooth particle 
hydrodynamics solver \citep{Pelupessy2005}. The mutual gravitational 
interaction between the gas particles and the stars are implemented 
using the BRIDGE solver \citep{Fujii2007}. BRIDGE provides a symplectic 
mapping for gravitational evolution in cases where the dynamics of a 
system can be split into two (or more) distinct regimes. In the 
application presented here the internal dynamics of the binary evolves 
on a relatively short timescale compared to the dynamics of the 
circumbinary disk. We adopt the bridge scheme to couple the 
gravitational dynamics of the inner binary with the hydrodynamics and 
self gravity of the circumbinary disk. The actual mutual forces are 
calculated by direct summation, the self gravity of the disc is 
calculated using a Barnes-Hut tree \citep {Barnes1986}. The gas is 
evolved with an isothermal equation of state.  We use the standard 
SPH viscosity formulation \citep{Monaghan1983} with $\alpha=0.5$ and $\beta=1.$ 
resulting in a Reynolds number of $R \sim 3000$.
 Additionally, the stars 
can act as sink particles to the gas. If a gas particles passes closer 
than 0.05 AU by a star it is accreted onto that star, transferring its 
mass and momentum to the star. 

\section{Simulations}
\label{Sec:sims}

\subsection{Initial conditions}
\label{Subsec:Initialconditions}

\begin{table}
\centering
\begin{minipage}{140mm}
\caption{Parameters of the initial conditions and planetary parameters \citep{Doyle2011,Welsh2012} 
for the Kepler runs.  Listed
are the binary semi-major axis ($a$), period ($P$), eccentricity ($\epsilon$), star masses and mass ratio 
($M_1$, $M_2$ and $q$), the total disk mass of the simulations ($M_d$) and the observed planetary 
semi-major axis ($a_{\rm planet}$), ratio of planet to binary semi-major axis
($a_{\rm planet}/a$) and eccentricity 
($\epsilon_{\rm planet}$). 
}
\label{table_ic}
\begin{tabular}{ l | c  c  c  c  c  c  c  c  c  c }

system         & $a$    & $P$      & $\epsilon$ & $M_1$     & $M_2$     & $q$    & $M_d$ & $a_{\rm planet}$ & $a_{\rm planet}/a$ & $\epsilon_{\rm planet}$    \\ 
               & (AU) & (days) &            & ($\MSun$) & ($\MSun$) &        & ($\MSun$) & (AU) &  & \\
\hline
\hline
Kepler 16      & 0.224 & 41.1     & 0.159       & 0.69      &  0.20     & 0.29   &  0.009  & 0.705 & 3.12 & 0.0068  \\
Kepler 34      & 0.228 & 27.8     & 0.521       & 1.05      &  1.02     & 0.97   &  0.021  & 1.09  & 4.76 & 0.182 \\
Kepler 35      & 0.176 & 20.7     & 0.142       & 0.89      &  0.81     & 0.93   &  0.017  & 0.603 & 3.43 & 0.042 \\
\hline
\hline

\end{tabular}
\end{minipage}
\end{table}

The initial conditions of the models consist of a binary star embedded 
in a circumbinary disk. For the parameters of the binary star we adopt 
the observed stellar masses, semi-major axis and eccentricity of Kepler 
16, 34 and 35 systems. In addition to this we also run models for 9 
known eclipsing binary stars taken from the \cite{Devor2008} catalogue 
to examine a wider range of parameters. We will discuss the latter runs 
in Section~\ref{sec:add}. 

The initial protoplanetary disk is set-up as an axi-symmetric disc with 
mass  $M_d$ constructed to be in equilibrium with a central 
mass $M_{tot}=M_1+M_2$. The disk mass adopted is $M_d=0.01 M_{tot}$.  The disk is assumed 
to be aligned with the binary orbit, which is plausible given the alignment of the binary and planetary 
orbits (within $0.5^\circ$). The 
discs have an inner boundary at $R_{in}= 2.4 a$ and an outer boundary 
$R_{out}=36 a$ (0.5 AU and 8 AU for the Kepler 16 model). The inner 
boundaries were chosen such that a limited number of gas particles are 
accreted onto the stars during the initial stages of the simulation. In 
this way the results do not depend on spurious accretion, as any smaller 
inner boundary would see the additional mass quickly accreted onto the 
stars. The density profile of the disk is $\Sigma(R) \propto R^{-\gamma}$,
with $\gamma=1$  and temperature profile $T(R) \propto 
R^{-3/4}$, normalized such that the Toomre Q parameter is equal to $Q=12$
at the disk edge, resulting in $T \approx 300$ K at 1 AU (for the Kepler 16 
model). The parameters of the binaries and discs for the Kepler 16, 34 and 35
systems are listed in Table 1. 

The Kepler 16, 34 and 35 systems turn out to sample different 
combinations of mass ratio q and eccentricity $\epsilon$, where Kepler 
34 and 35 have $q \approx 1$ and Kepler 16 has $q \approx 0.3$ (see Table 
\ref{table_ic}). Kepler 16 and 35 have similar low $\epsilon \approx 0.15$ 
while Kepler 34 has a high eccentricity $\epsilon=0.52$. 

\subsection{Results}
\label{Subsec:runs}

\begin{figure*}
 \centering
 \epsfig{file=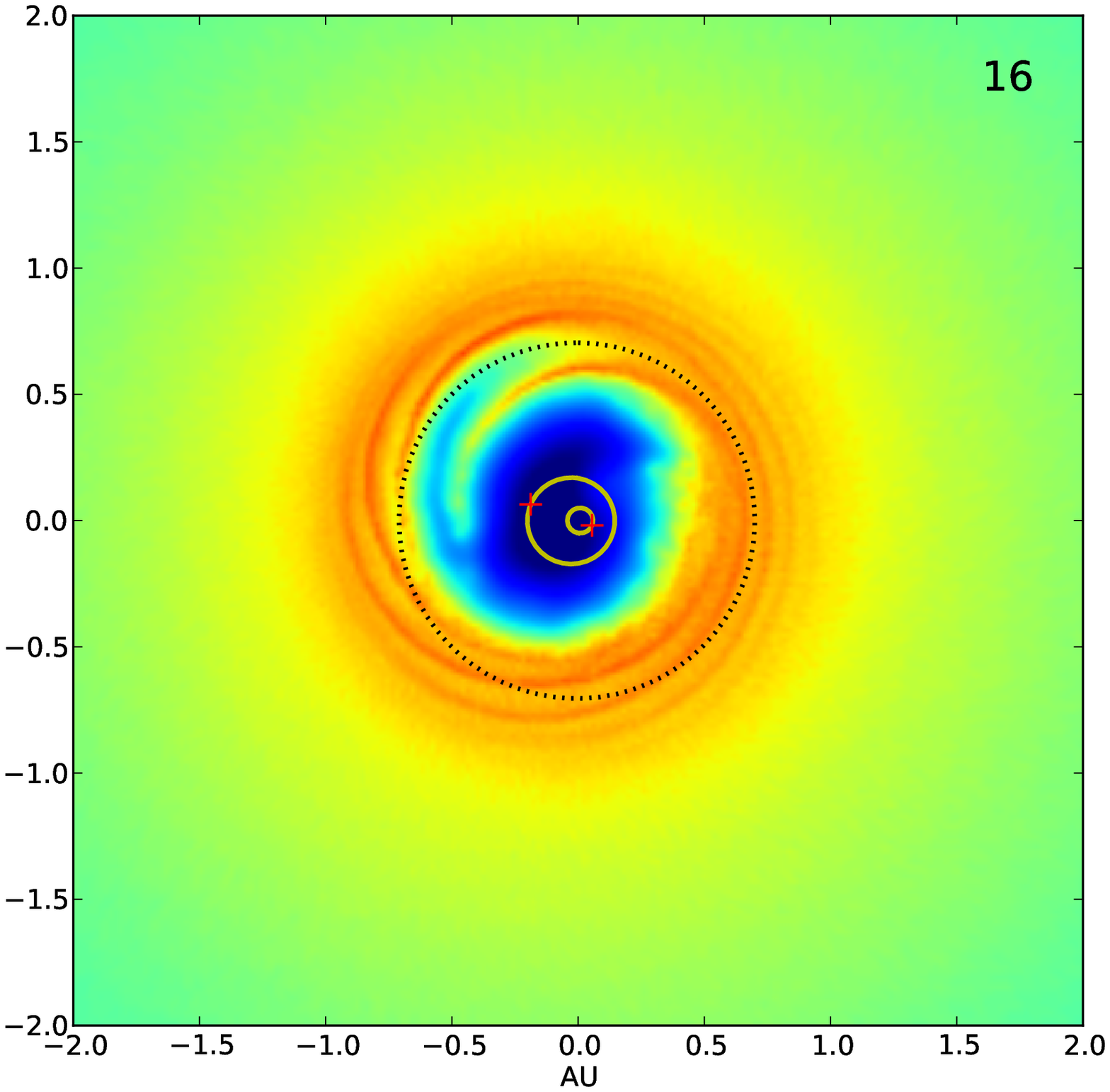, width=.42\textwidth}
 \epsfig{file=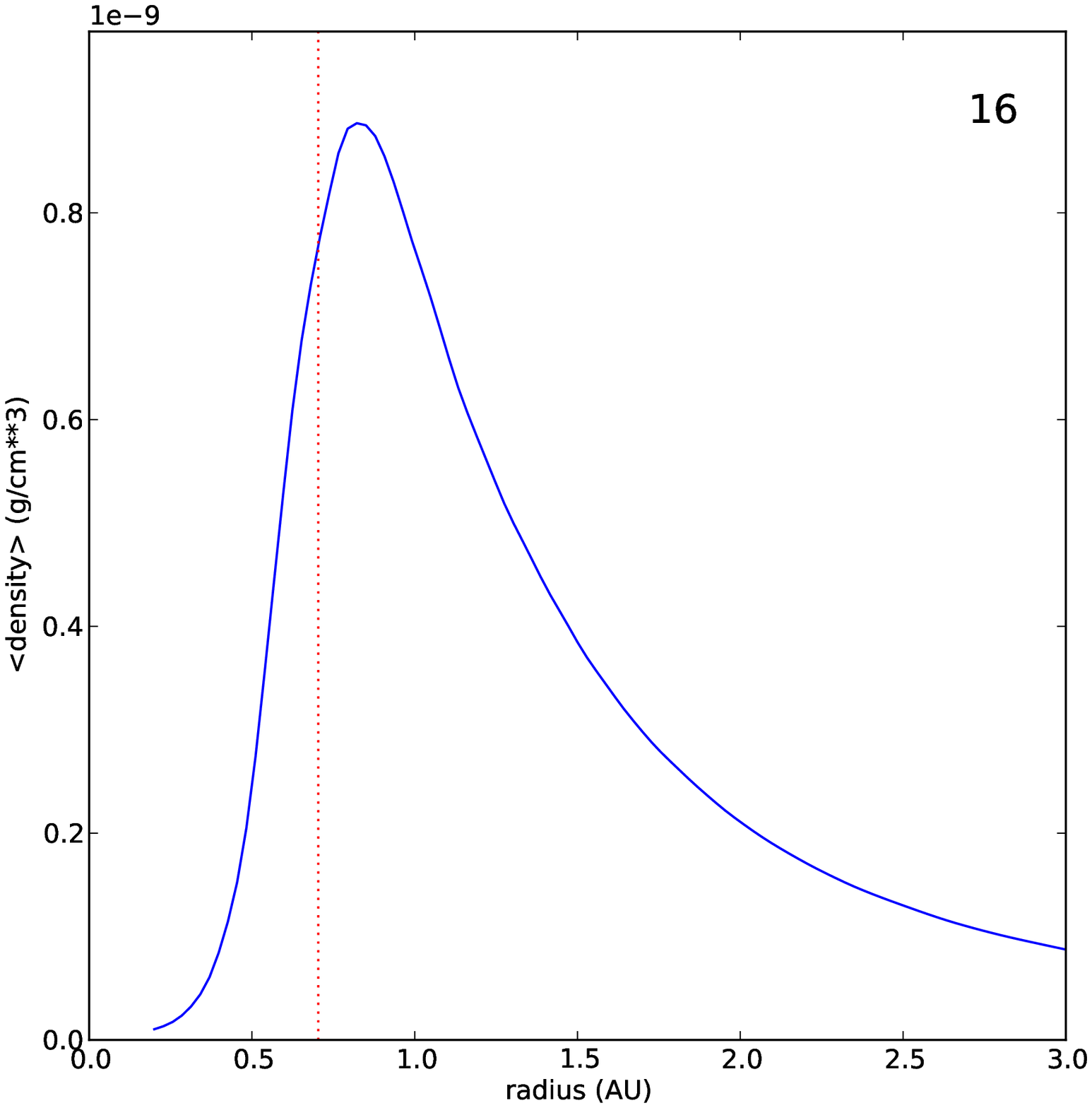, width=.42\textwidth}

 \epsfig{file=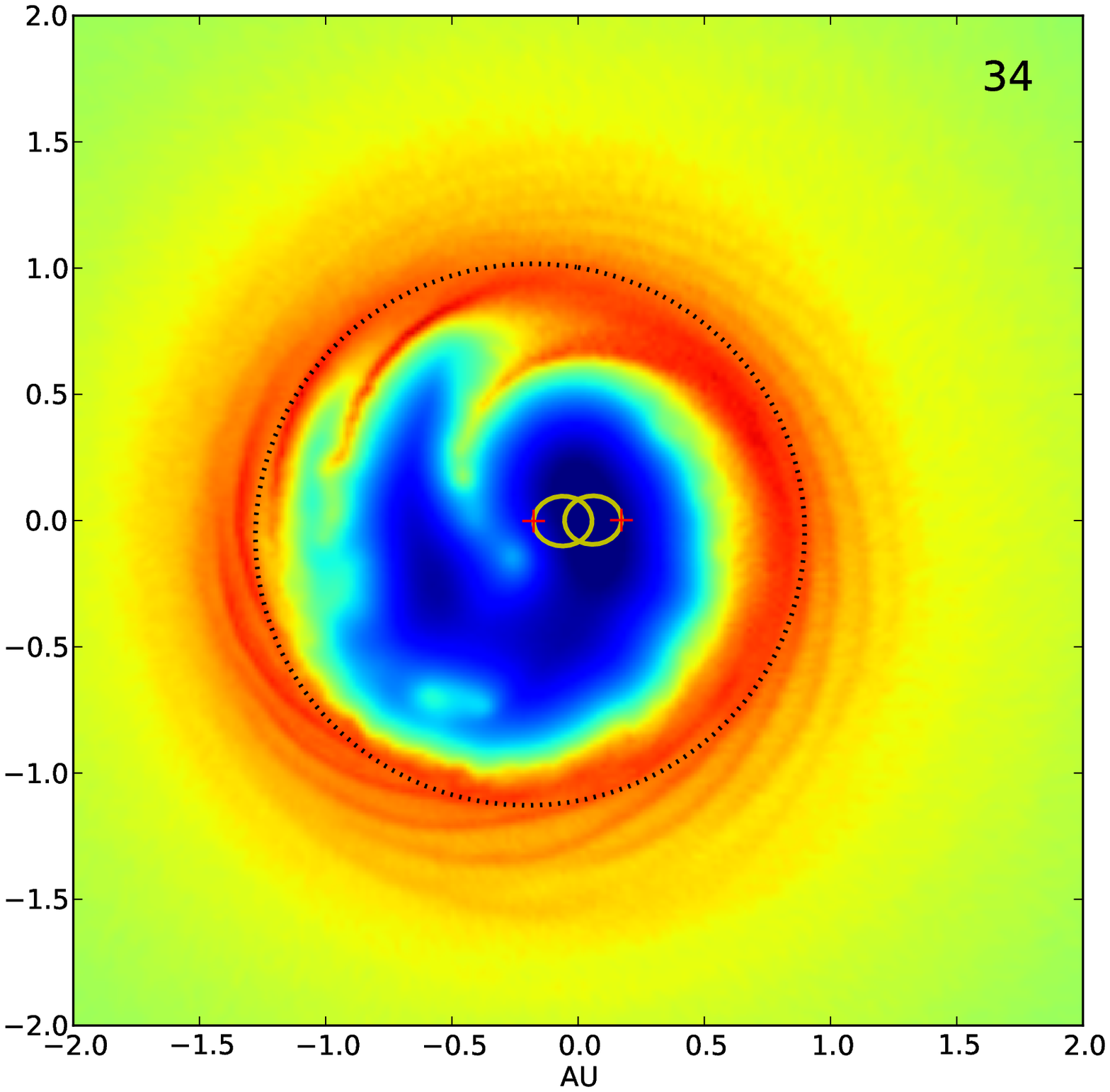, width=.42\textwidth}
 \epsfig{file=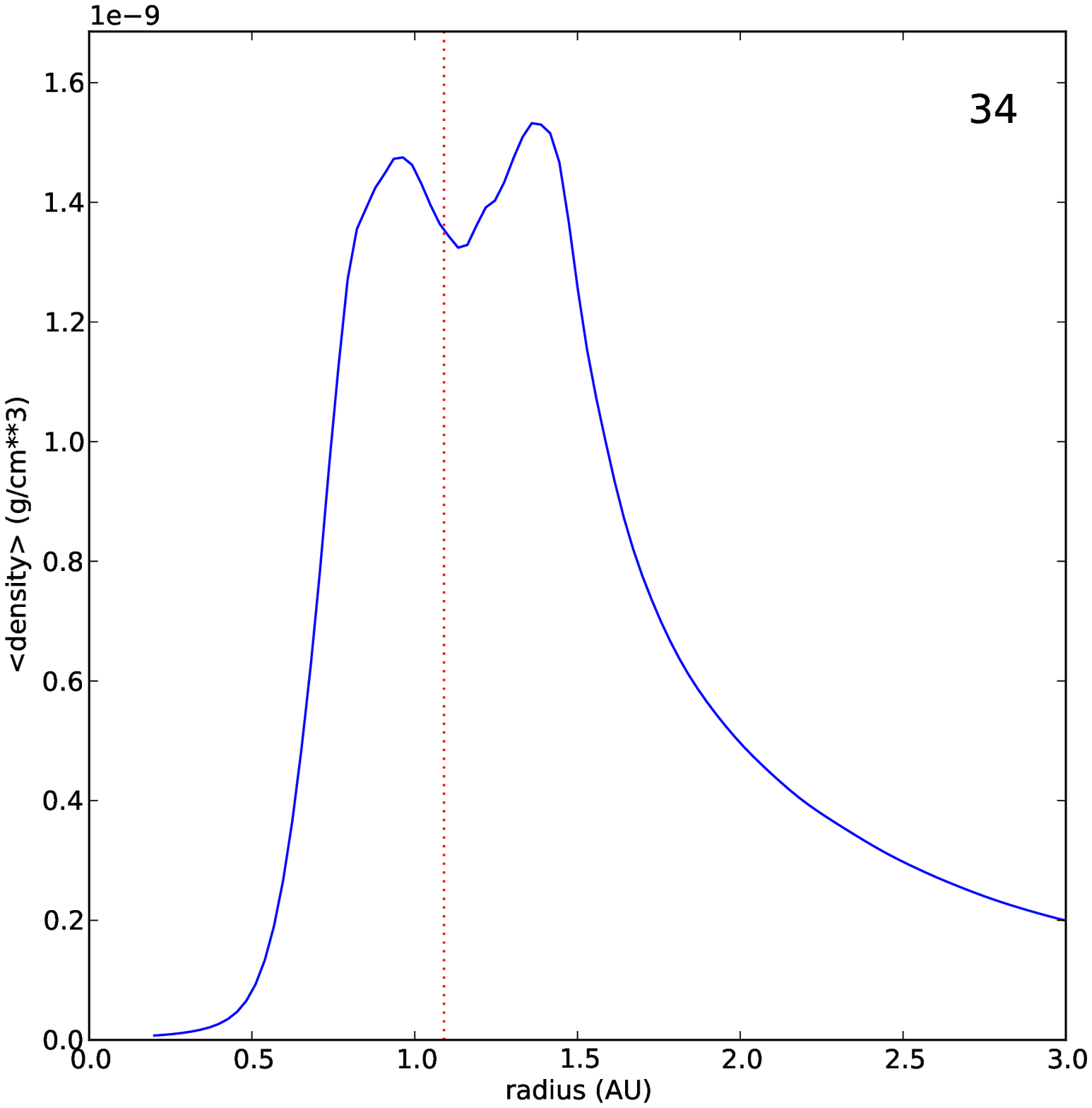, width=.42\textwidth}

 \epsfig{file=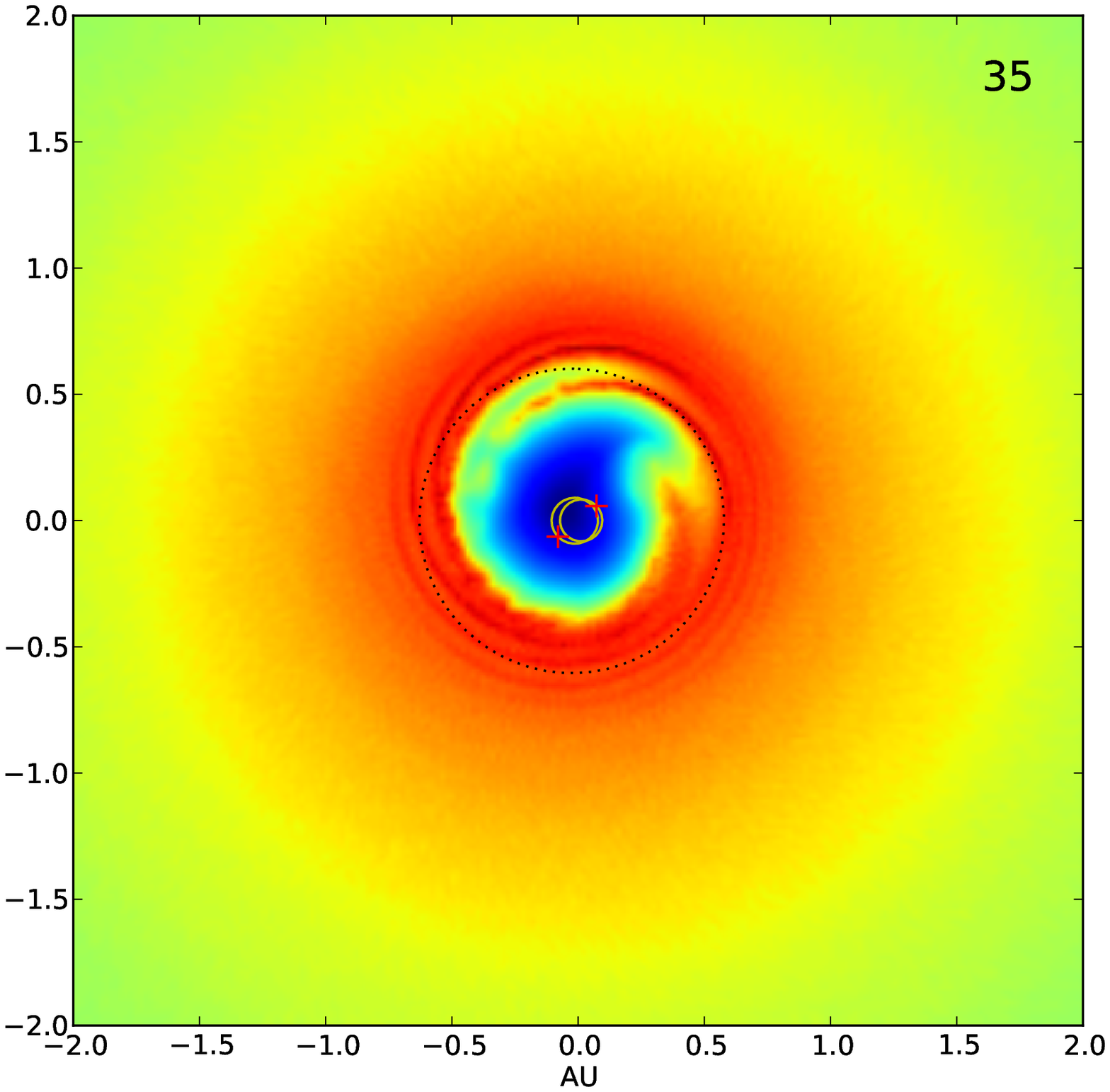, width=.42\textwidth}
 \epsfig{file=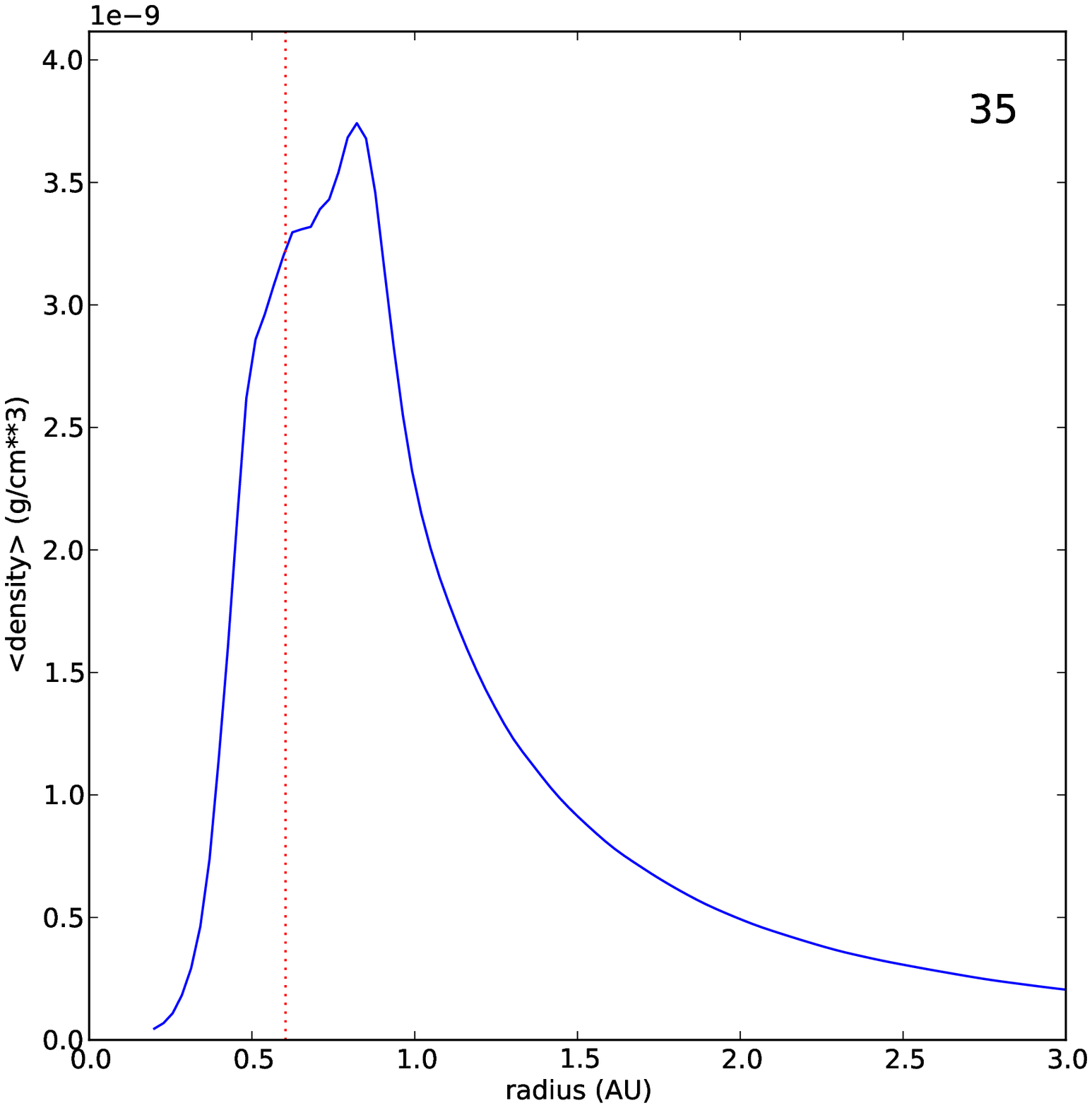, width=.42\textwidth}

 \caption{Left panels: slices of gas density. Plotted is a slice through 
 the midplane gas density. Also plotted are the orbit of the binary and 
 the approximate orbit of the observed planets (the periapsis angle is 
 matched to the average periapsis angle of the disc material). Right 
 panels: azimuthally and time averaged midplane density vs radius. 
 Dotted line indicates the radius where the planet in the respective 
 systems is found.}
 \label{fig:xy} 
\end{figure*}

We run the simulations for 1000 binary orbits. A snapshot of the
resulting gas distribution after approximately 300 orbits for the
Kepler 16, 34 and 35 runs are plotted in figure \ref {fig:xy}. At this
point the distribution is in quasi-stationary equilibrium where the
distribution of gas changes little qualitatively afterwards.  Compared
with the initial distribution of gas, the inner disc gap has expanded,
and an eccentric overdense ring has formed \cite[in agreement
  with][]{Artymowicz1994, Artymowicz1996}. The gas disk has been
pumped into an eccentric orbit by high order Lindblad resonances
\citep{Pierens2007}. The interaction is highly non-linear though and
the inside of the gas flow exhibits strong periodic tidal
streams. Close inspection of the time lapses of the simulation shows
that the onset of these tidal streams coincides with the start of the
eccentricity pumping.  The action of the resonances excites strong
waves in the gas distribution.  The azimuthally averaged density plots
in Figure~\ref{fig:xy} (right panels) are for this reason time
averaged over 10 snapshots (spanning the 100 last
orbits).  For all three systems there is a close correspondence of the
location of the dense ring and the planet.

 In order to test the robustness of our main conclusions we varied 
the parameters of the Kepler 16 run to test the effects of (small) disk 
inclinations ($i=5^\circ$),  variations in Toomre Q parameter ($Q=8,16$), $\gamma$ ($\gamma=1.5$) and 
disk mass fraction ($M_d/M_{tot}=0.02$), as well as numerical resolution, 
running at $N=10^5$ and $N=10^6$. The results presented here are insensitive 
to these parameter variations.

\begin{figure*}
 \centering
 \epsfig{file=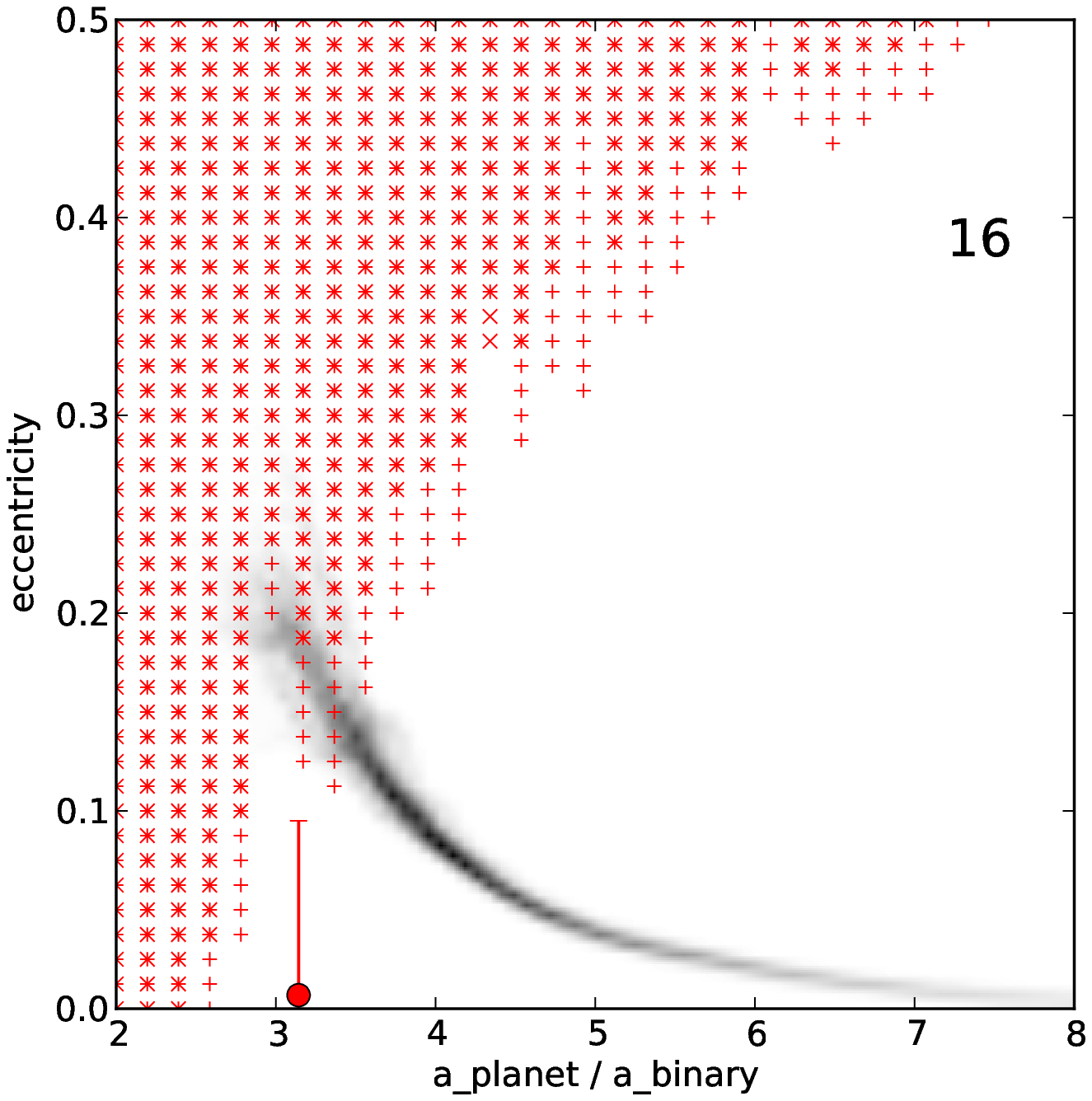, width=.42\textwidth}
 \epsfig{file=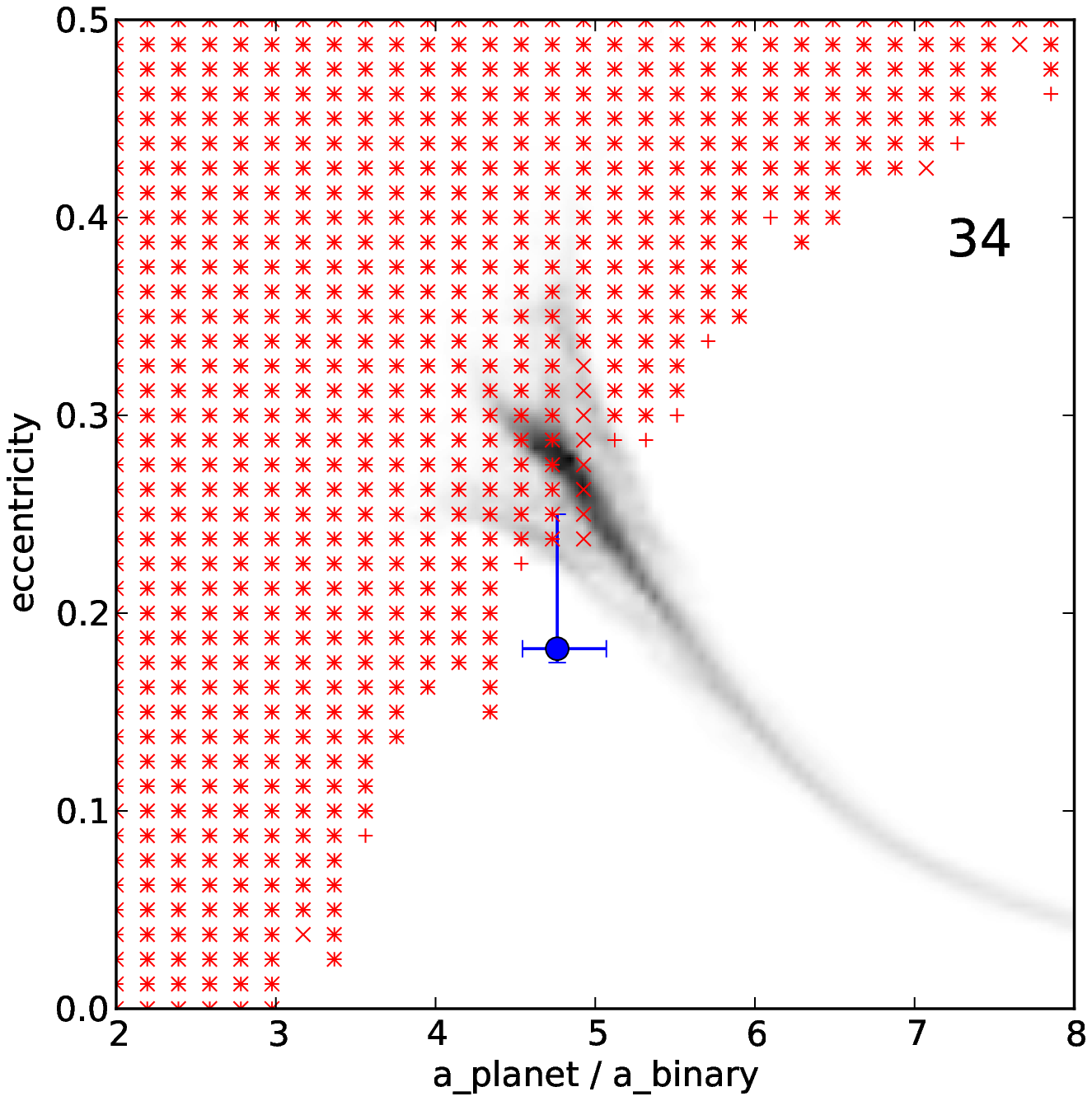, width=.42\textwidth}
 \epsfig{file=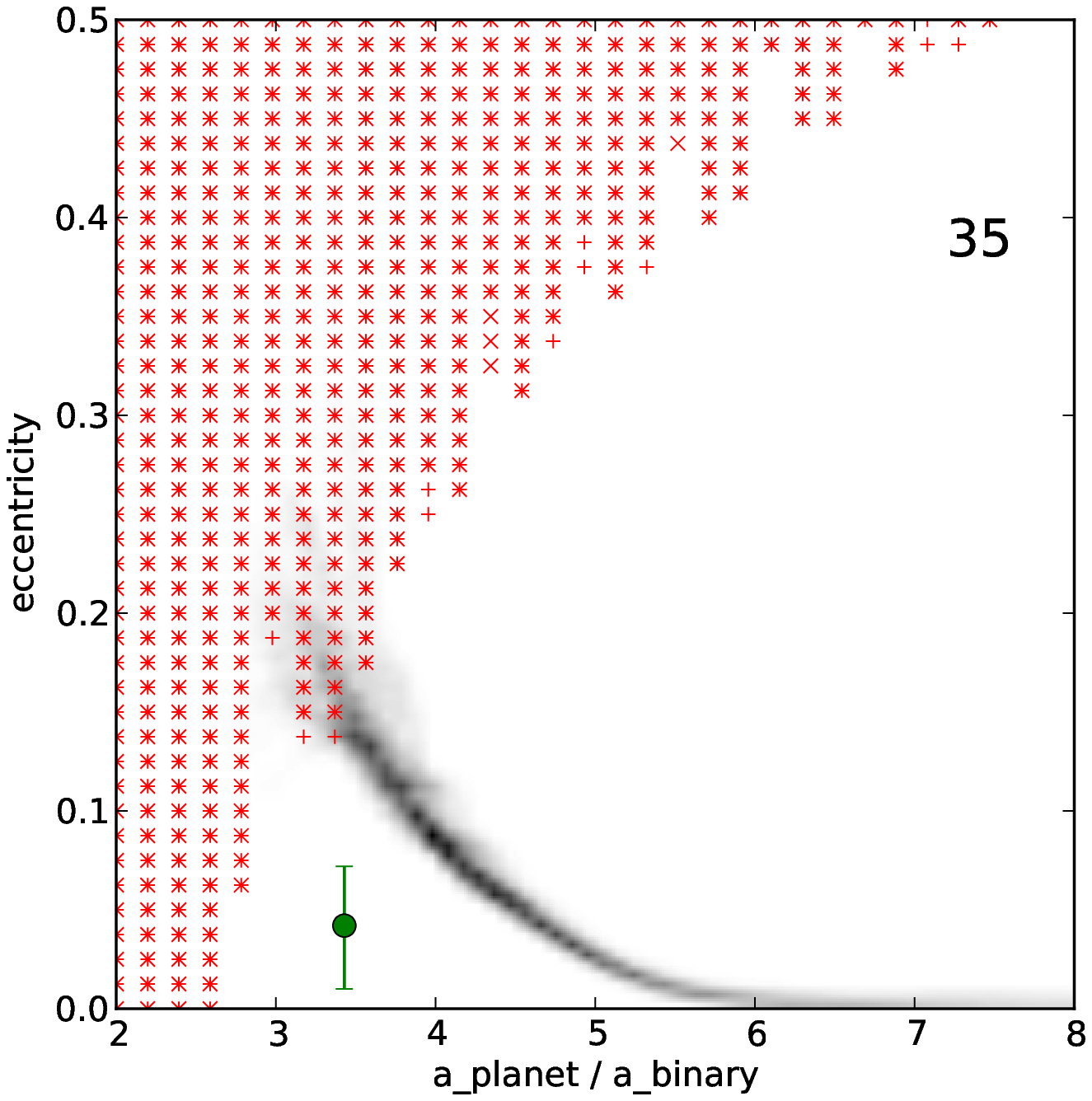, width=.42\textwidth}

 \caption{ Semi-major axis - eccentricity distribution of the gas.
   Plotted is the density weighted phase space density in $a$ and $e$
   of the disc material (From top to bottom the models are: 16,34 and
   35).  Semi major axis is scaled with the binary semi-major axes of
   the respective system. The points with bars in each panel indicate
   the  observed orbital parameters  of each system.  The bars indicate the
   range in eccentricities (and for Kepler 34 also semi-major axes)
   encountered in long term integrations of the triple binary$+$planet
   systems. The red crosses and plusses show the result of a grid of
   binary$+$planet integrations with different $a$ and $\epsilon$, and pericenter
   angle $\alpha_p$, where plusses are unstable systems for $\alpha_p=0$ and 
   crosses unstable systems for $\alpha_p=\pi$.}
 \label{fig:aeps}
\end{figure*}

Figure \ref{fig:aeps} shows the distribution of the (instantaneous)
semi major axis $a$ and eccentricity $\epsilon$ of the disk material
for the three runs. As can be seen the gas follows a relatively narrow
relation in the $a-\epsilon$ plane where the material in the inner
disk follows increasingly eccentric orbits, while the outer disc
material follows circular orbits. In these figures the points with
bars indicate the position of the observed planets in the $a-\epsilon$
diagram. The bars do not indicate the uncertainty in the observations
(these are very small) but indicate the secular variation of the orbital
elements in time, derived from long-term three body integrations
\citep{Doyle2011, Welsh2012}. For Kepler 16, most of the variation consists
of slow (on a timescale of $\approx 40$ yr) oscillations around the forced eccentricity
due to secular pertubations from the binary, while for Kepler 34 and 35 most 
of the variation in the eccentricity is due to fast oscillations on the orbital timescale
\citep[as expected for equal mass binaries;][]{Moriwaki2004}.
For all models the eccentricities in the
gas disk is greater than the observed planetary orbit even accounting for
the variations in the orbital elements, although for Kepler 34 the disk
material falls only just outside the range of the long-term
integrations. Although we only have three systems, and the
eccentricities are poorly matched, the system with the larger disc
eccentricity coincides with the larger planetary ecccentricities.

An interesting question is whether any planets forming in the disk
would be stable (if they inherited the orbital elements from the
disk). In order to investigate this we have run a grid of three body
models for each of the Kepler binaries with the central binary and
semi-major axis $a_{\rm planet} = 2-10 a_{\rm binary}$ and $\epsilon=0-0.5$
choosing a grid of 40x40 models. 
For the simulations we employed the Huayno
solver \citep{Pelupessy2012} in the AMUSE package, using its
order 10 shared variable timestep (SHARED10) integrator
\citep{Sofroniou2005}. We ran models at each $a$ starting from
$\epsilon=0.5$ and going to lower $\epsilon$ until a stable model was
encountered (a model was deemed unstable if a deviation of $50\%$ in
$a_{\rm planet}$ was encountered - models were run for 1 Myr).  Considering coplanar
orbits the most important additional parameter is the angle $\alpha_p$ between binary and
planetary pericenter. For a given $a$ there is a small range in $\epsilon$ where a model 
can be stable or unstable depending on the pericenter angle. The
results are summarized in figure \ref{fig:aeps},  where models which are always unstable
are marked with stars, while models which can be stable 
depending on pericenter angle are marked with plusses or crosses. In the case of Kepler 16, 
varying the pericenter angle shows the biggest change in the unstable $\epsilon$, because of 
the larger mass ratio between the binary components. Note that all three observed
planets lie within the stable region. Also, the disc material
at the radius of the planet is in the unstable  or partially stable region for 
all the models. This makes a scenario whereby the disc material is converted
in situ into a planet (inhereting both semi-major axis and
eccentricity,  assuming that the dissolution of the gas disk
does not alter the orbits of the remaining planetesimal material) 
followed by slow evolution of the orbital elements
unlikely, unless inward radial migration has occurred.

\subsection{Additional models}
\label{sec:add}

In order to examine a wider variety of parameters we run in the same way 
a set of models varying the mass and eccentricity of the  primary stars, 
which we chose from a catalog of 773 eclipsing binaries \citep{Devor2008}. 
The systems were selected from the \cite{Devor2008} survey 
on the basis of their orbital period ($P>10$ days) and eccentricity 
($\epsilon>0$).  This ensures that we select systems that are unlikely to 
have undergone tidal interactions, in which case they are unsuitable
to our analysis. 
This results in a set consisting of 9 systems  (note this does not exclude 
the possibility of planets around the remaining 764 systems). An overview of the 
initial conditions of this ``survey run'' is given in Table~\ref
{table_survey}.

\begin{figure}
 \centering
 \epsfig{file=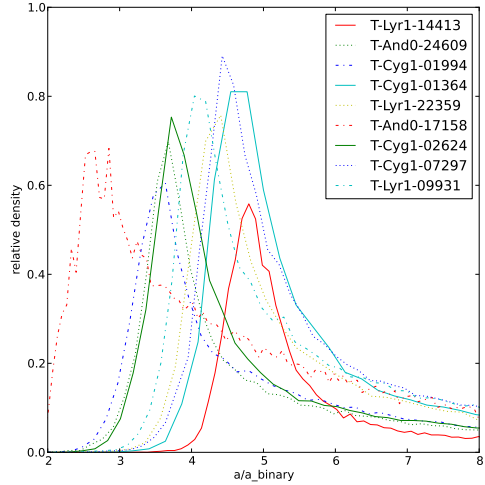, width=.45\textwidth}

 \caption{Distribution of gas vs semi-major axis for the survey runs.}
 \label{fig:adist}
\end{figure}

The resulting gas disks from these runs were analyzed in the same way as 
in Sect.~\ref{Subsec:runs}, i.e. we calculate the semi-major axis and 
eccentricity of the disk material. In Figure~\ref{fig:adist} we plot the 
resulting density distributions as a function of semi-major axis. 
 If planets commonly form around close binaries we could expect 
them to form preferentially close to the peak of the density 
distribution with an eccentricity smaller than that of the disk material, 
as they have been seen for the hitherto detected circumbinary Kepler systems.
If this is indeed the case we can derive the expected location of any planets 
in these systems from the simulations. We present in Table~\ref{table_survey} 
fiducial planetary elements $a_{\rm planet}$ and $\epsilon^{\rm up}_{\rm 
planet}$. The eccentricity could lie anywhere between 0 and 
$\epsilon^{\rm up}_{\rm planet}$, whereas the semi-major axis would be 
expected to lie within $\approx 20\%$ of the $a_{\rm planet}$ given.

\begin{figure}
 \centering
 \epsfig{file=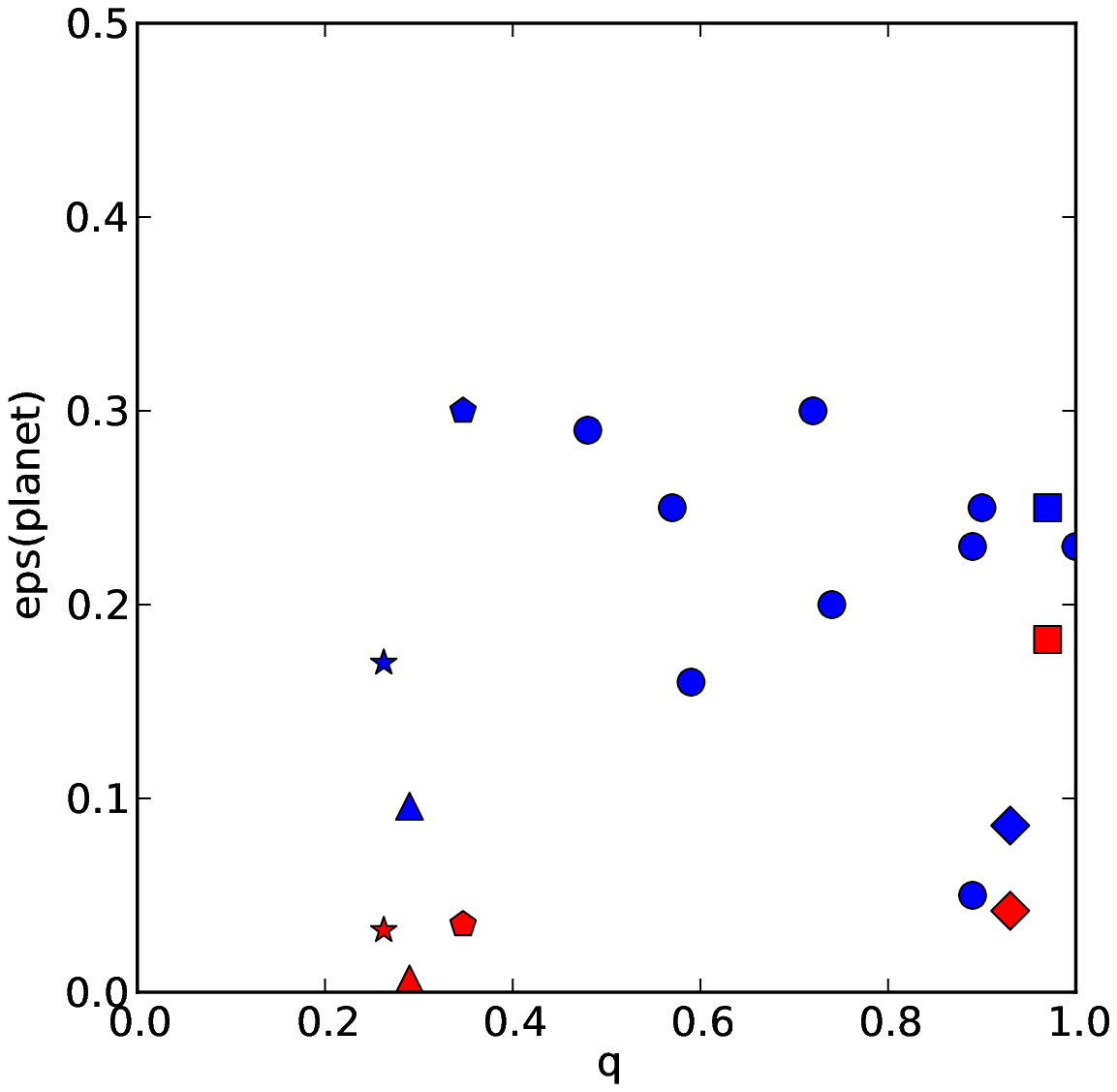, width=.42\textwidth}
 \epsfig{file=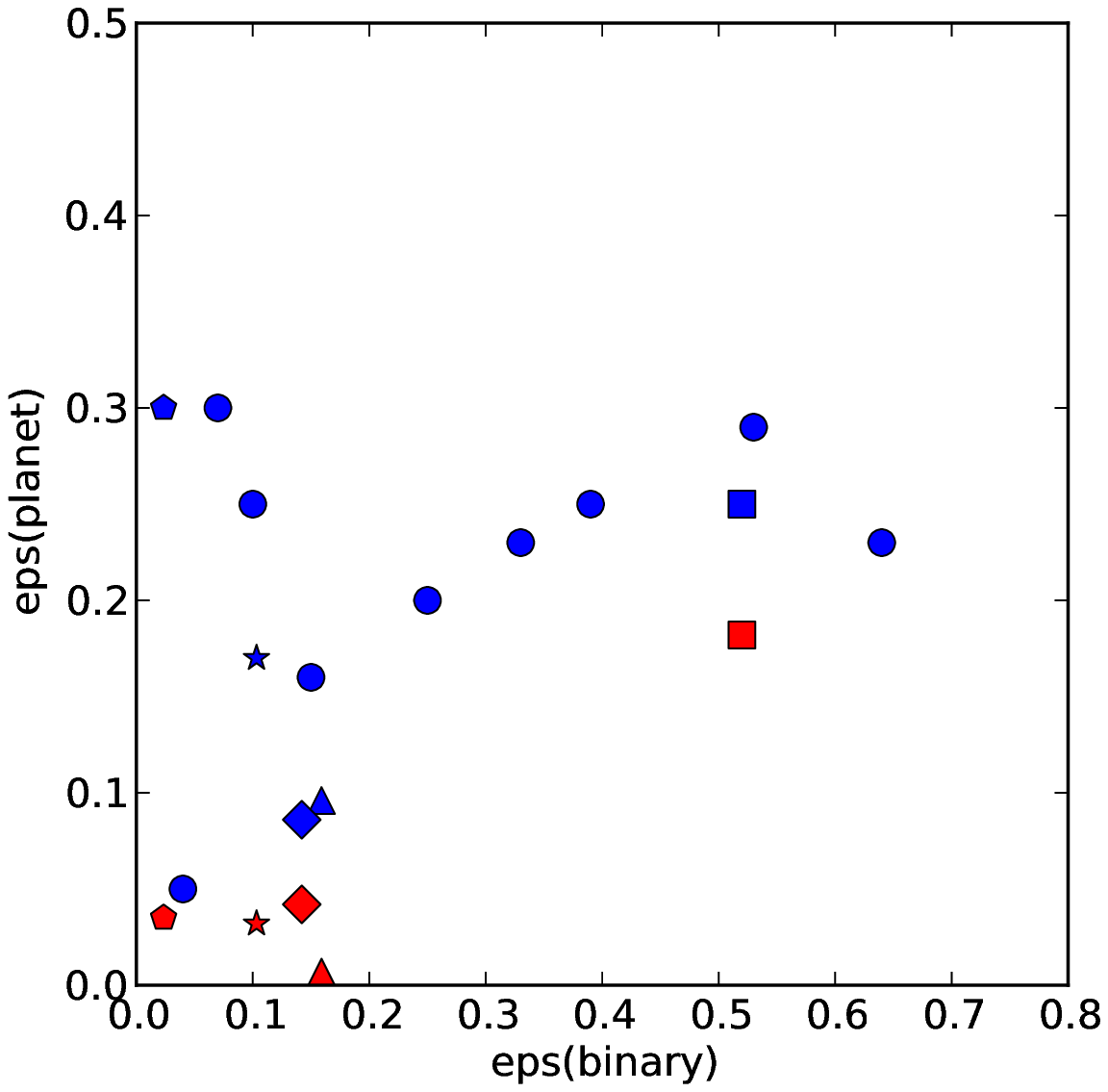, width=.42\textwidth}

 \epsfig{file=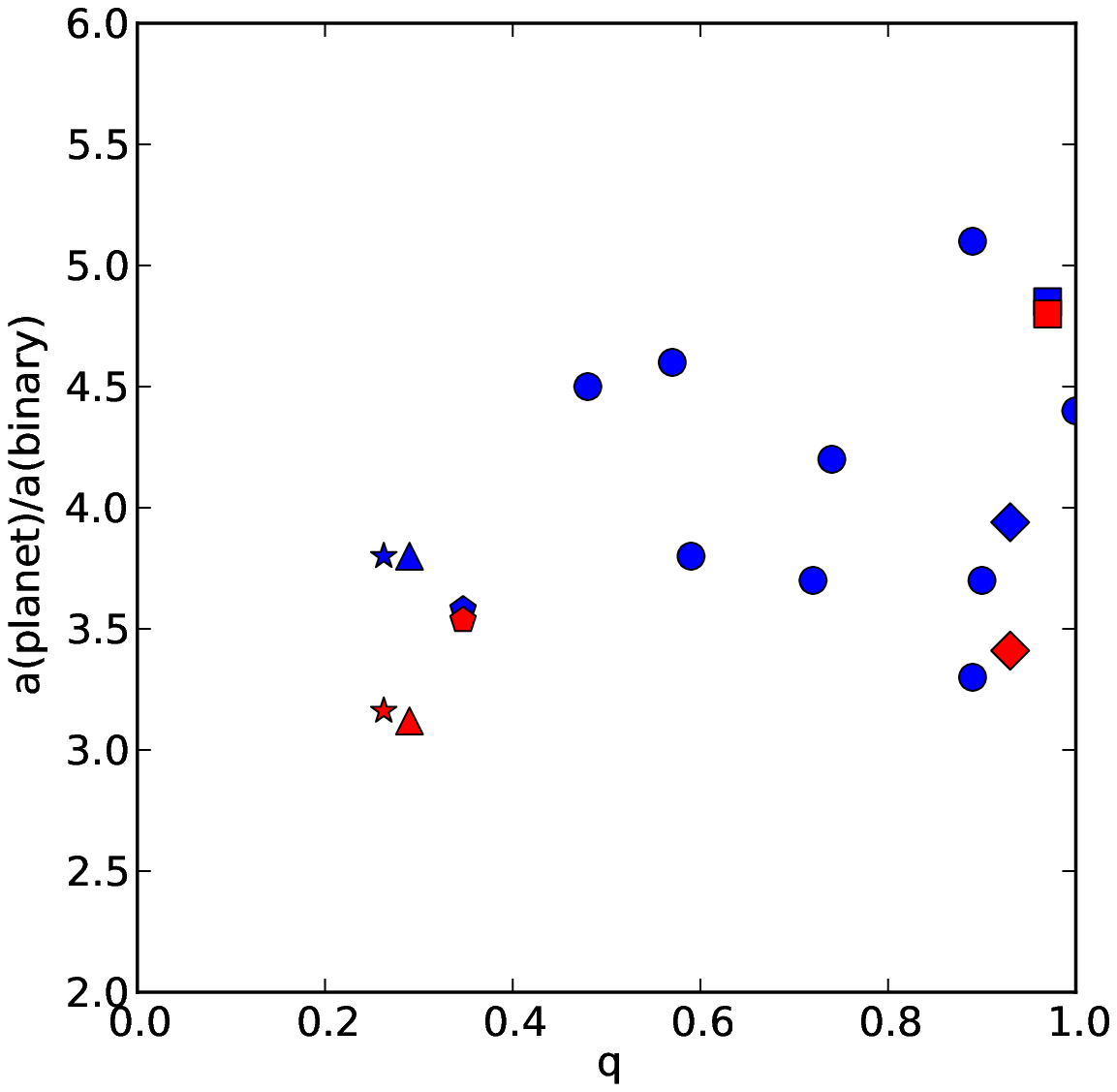, width=.42\textwidth}
 \epsfig{file=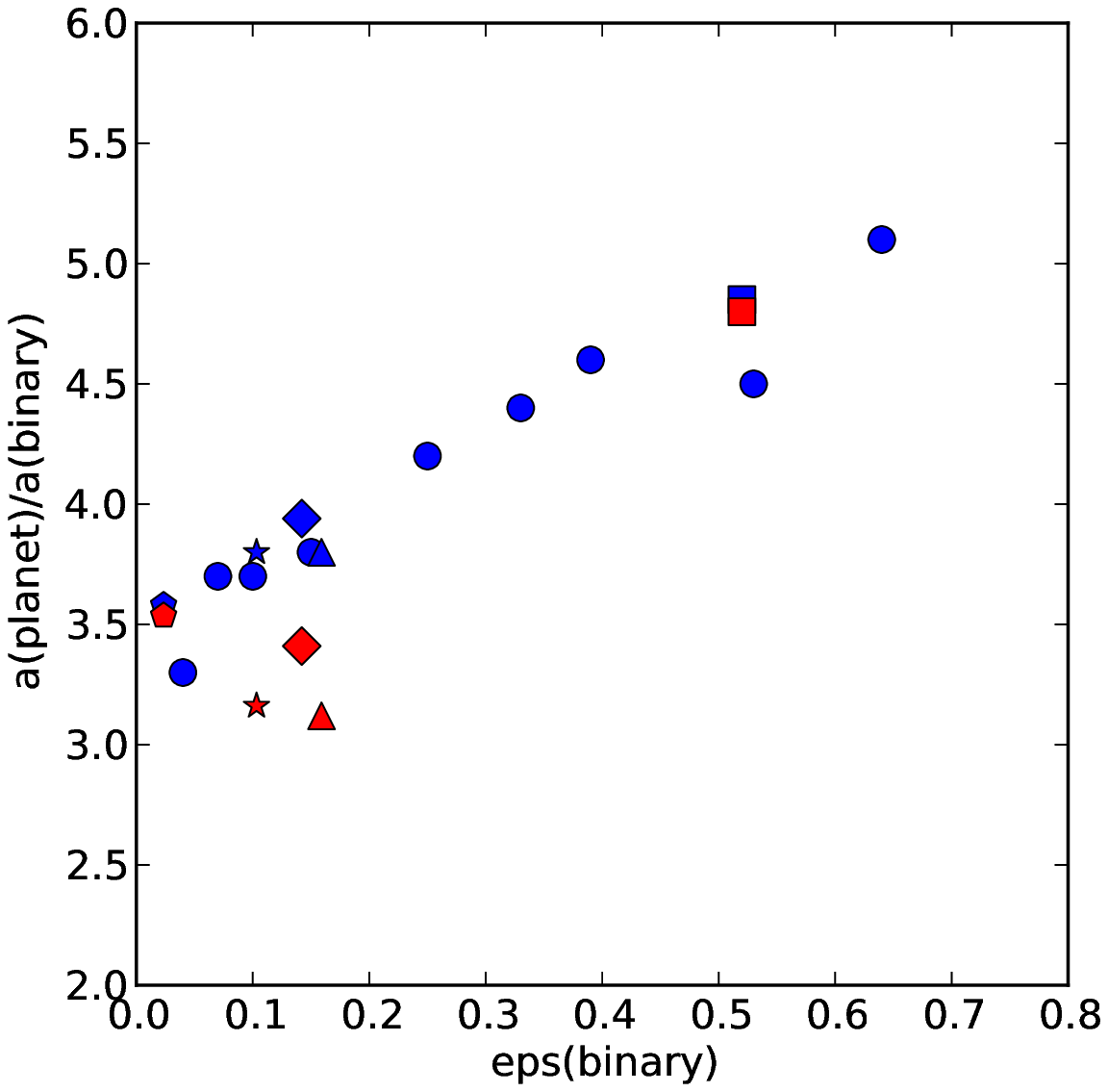, width=.42\textwidth}
 
 \caption{Scatter plots of binary vs planet or disk properties. The red 
 symbols are the observed Kepler systems  (triangle, square, diamond, star and pentagon 
 symbols for Kepler 16,34,35,38 and 47 (inner planet)), while the blue symbols are 
 results from the simulations  (matching symbols for the Kepler systems,
 circles are the survey systems from Table~\ref{table_survey}), where we plot the binary 
 properties against the orbital elements of the peak in the density of 
 the semi-major axis - eccentricity distribution of the disk material. 
 Plotted are the binary mass ratio ($q$) against eccentricity of the 
 planet or disk (upper left panel), binary eccentricity against planet 
 eccentricity (upper right), binary mass ratio against planet semi-major 
 axis (in units of the binary semi-major axis, lower left) and the 
 binary eccentricity against the planet semi-major axis(lower right).}
 
 \label{fig:corr}
\end{figure}

In Figure~\ref{fig:corr} we show a number of scatter plots of binary 
parameters (mass ratio $q$ and eccentricity $\epsilon$) versus 
`planetary' orbit semi-major axis $a_{\rm planet}$ and eccentricity 
$\epsilon_{\rm planet}$. For comparison we include the observed planets 
Kepler 16, 34, 35, 38 and 47 \footnote{Note Kepler 47 is actually a two planet system, we consider
the parameters of the inner planet.}. $a_{\rm planet}$ nor $\epsilon_{\rm planet}$ show 
a correlation with the binary mass ratio $q$, and neither does the  
$\epsilon_{\rm planet}$-$\epsilon_{\rm binary}$ plot. However, we see 
that the ratio of planetary-to-binary semi-major axis $a_{\rm 
planet}/a_{\rm binary}$ follows a tight relation with the binary 
eccentricity. The observed systems fall on the same relation. We should 
be careful interpreting the (absence of) $\epsilon_{\rm planet}$ correlations
as these are only upper limits. However the $a_{\rm planet}/a_{\rm binary}$ correlation, 
 which is basically the same as the relation of disk gap size and eccentricity 
found by \cite{Artymowicz1994} since the surface density rises quickly to a peak 
at the inner disk edge,
agrees with a calculation of the non-intersecting invariant orbital 
loops in binary systems by \cite{Pichardo2005}, who found that the radius of 
the inner non-intersecting orbit does not depend strongly on $q$, but 
increases with $\epsilon$. The correlation between binary
eccentricity and planet-binary semi-major axis ratio we find 
\begin{equation}
  a_{planet} \simeq \left(3.2 + 2.8\,e_{\rm binary}\right) a_{\rm binary}
\end{equation}
which is expected in case circumbinary planets form preferentially at 
the peak density in the disk, should be testable as  a more statistically 
significant number of planets around binary systems are found,  especially as so
far only one system with binary eccentricity $\epsilon>0.2$ is found.

\begin{table*}
  \centering
  \begin{minipage}{140mm}
  \caption{Initial conditions and predictions for planetary orbits for 
  survey systems. Same initial conditions as in Table~\ref{table_ic} are 
  given, and additional the results for the semi-major axis and a (loose) 
  upper limit (see text for discussion) for the eccentricity of the 
  fiducial planet orbits (defined by the peak in the semi-major axis - 
  eccentricity distribution of the disk material). }
  \label{table_survey}
  \begin{tabular}{ l | c  c  c  c  c  c  c | c c}
  
  system         & a    & P      & $\epsilon$ & $M_1$     & $M_2$     & $q$    & $M_d$   & $a_{\rm planet}$ & $\epsilon^{\rm up}_{\rm planet}$  \\ 
                 & (AU) & (days) &            & ($\MSun$) & ($\MSun$) &        & ($\MSun$) & (AU) &  \\
  \hline
  \hline
  T-Lyr1-14413   & 0.29 & 39.9 & 0.64 & 1.08 & 0.96 & 0.89 & 0.020 & 1.47 & 0.23 \\
  T-And0-24609   & 0.18 & 18.0 & 0.10 & 1.22 & 1.10 & 0.90 & 0.023 & 0.65 & 0.25 \\
  T-Cyg1-01994   & 0.17 & 14.5 & 0.15 & 1.80 & 1.06 & 0.59 & 0.029 & 0.63 & 0.16 \\ 
  T-Cyg1-01364   & 0.12 & 12.2 & 0.53 & 1.03 & 0.50 & 0.48 & 0.015 & 0.50 & 0.29 \\
  T-Lyr1-22359   & 0.13 & 12.3 & 0.33 & 0.97 & 0.97 & 1.00 & 0.019 & 0.57 & 0.23 \\
  T-And0-17158   & 0.12 & 11.4 & 0.04 & 1.03 & 0.92 & 0.89 & 0.020 & 0.41 & 0.07 \\
  T-Cyg1-02624   & 0.15 & 11.6 & 0.07 & 2.11 & 1.52 & 0.72 & 0.036 & 0.57 & 0.30 \\
  T-Cyg1-07297   & 0.12 & 11.6 & 0.39 & 0.97 & 0.55 & 0.57 & 0.015 & 0.52 & 0.25 \\ 
  T-Lyr1-09931   & 0.12 & 11.6 & 0.25 & 0.91 & 0.67 & 0.74 & 0.016 & 0.49 & 0.20 \\
  \hline
  \hline
  
  \end{tabular}
  \end{minipage}
\end{table*}

\section{Discussion}
\label{Sec:Discussion}

The simulations presented here favor the hypothesis that the structure
of a circumbinary disc and the planetary orbit of the observed
systems Kepler 16,34 and 35 are closely related. The agreement in the
correlation between binary eccentricity and planet semi-major axis
shown in Figure~ \ref{fig:corr} of the observed and simulated systems
favour this simple scenario.

In principle there are three possibilities to account for such a 
correspondence: 1) The planet has formed in situ in the overdense 
region, 2) the planet has formed further out in the disk and migrated 
inward while the circumbinary disc was still present until it 
encountered the region of positive density slope at the gap ( in this case the
the planet occupies an orbit close to the peak density just because the 
surface density rises quickly at the inner edge), and 3) the planet 
has formed further out in the disc and migrated by secular evolution 
(e.g. scattering by planetesimals) and stalled at the radius 
previously occupied by the inner disk edge for reasons that are not 
directly related to the previously existing gas disk.

Upon naive inspection of figure \ref{fig:xy} alternative (1) seems 
preferable. However in this case the planet eccentricities are smaller 
than expected from figure~\ref{fig:aeps} and furthermore  planetesimals 
originating from the gas disk at the current semi-major axis of the planet 
would be in marginally stable orbits. 
This problem could be remedied by a mechanism whereby the 
eccentricity of a planet would decrease (e.g. by forming multiple 
proto-planets and having them collide). Given the fact that at the radii 
of the planets orbits of modestly higher eccentricities are already 
unstable, it seems that a modest amount of radial migration (i.e. moving
to  the left in the diagram of fig.~\ref{fig:aeps}) is preferred.   

 Scenario 3) - migration after the disc has cleared - suffers from 
the problem of fine-tuning the resulting orbit to match the location of 
peak density at the inner disc edge. In this respect 2) provides a more 
natural migration scenario. Additionally, gravitational scattering by 
larger planets (which may be supported by the detection of a binary 
system with multiple planets, Kepler 47) is unlikely given the 
fact that the eccentricities of most of the Kepler circumbinary planets are quite low. 
These planets are too far out for tidal circularisation (such as in the 
case of hot jupiters). 

Of course the above conclusions are based on a simplified model
where we have ignored the subsequent phases of planet formation (the
build up of planetesimals and their growth). We argue, based on the
close comparison, that these may not be essential in determining the
planetary orbit characteristics for the special case of circumbinary
planet formation. If indeed this is the case a more careful
consideration of the latter phases in the disk evolution where the
remnant material is cleared out while the planet emerges from the disc
may still be needed to show that the planets end up in their long-term
stable configuration.

\section{Conclusions}
\label{Sec:Conclusions}

We investigated the formation mechanism of the recently discovered 
population of planets orbiting both components of a binary by studying 
hydrodynamic simulations of circumbinary disks. We chose binary 
parameters according to the observed systems Kepler 16, 34 and 35 and an 
initial disk with canonical initial conditions. The disks were evolved 
until they settled in quasi equilibrium and the resulting systems were 
compared with the observed systems. We also run three-body simulations 
to investigate the long term stability of planetary orbits around the 
binaries. We performed 9 additional simulations for which the initial 
conditions are taken from a survey of eclipsing binaries.  In Tab.\,\ref
{table_survey} we present, for these observed binary systems, the 
predicted semi-major axis and eccentricity of the circum binary planet.  
We further conclude that:

\begin{itemize}
  \item Planets observed are found inside the theoretically expected 
        overdensity at the inner edge of the circumbinary disk,
  \item The material in the circumbinary disks itself settles in a narrow 
        range of eccentricities,
  \item The eccentricity ranges of the observed planets are smaller than 
        that of the disk material, with the possible exception of the Kepler
        34 model.
  \item A relatively tight positive relation between planet 
        semi-major axis and binary eccentricity is  expected if planets
        form preferentially at the density peak just outside the inner edge of 
        the circum binary disk.      
\end{itemize}

These results suggest that planet formation in these systems   and therefore
the orbital parameters of these planets, are 
determined for a large part by the binary driving of the proto-planetary 
gas disk. In addition it seems necessary that the systems we have modelled 
in detail, Kepler 16,34,35, have experienced at least some planetary 
migration. Formation in the outer parts of the disk followed by 
migration inward is possible, but the naive expectation that the planet 
would be found inside the disk gap is not entirely fulfilled. This 
formation mechanism would need some fine-tuning but not as severely as 
for planetary formation models around single stars (in so far as these 
invoke planetary trapping). A direct relation between disk material, 
e.g. through gravitational collapse, is probably not the formation 
route, as this would imply that the planets would form with an 
eccentricity close to the material of the disk. If this were the case 
planets with this eccentricity could in principle also form, 
since most of the high eccentricity disk material is well within the 
region of stability. Planetesimal formation from the ring material 
followed by aggregation is not excluded. In this scenario the 
eccentricity of the planet is lowered through repeated collisions. In 
this case the close correspondence between the gas ring in our 
simulations and the planets is not accidental. 

{\bf Acknowledgements}
This work was supported by the Netherlands Research Council NWO 
(grants \#643.200.503, \#639.073.803 and \#614.061.608) and by 
the Netherlands Research School for Astronomy (NOVA).

\bibliographystyle{mn2e}

\end{document}